\input{aipcheck}

\documentclass[draft]{aipproc}

\layoutstyle{6x9}

\begin{document}

\title{Hidden Quantum Markov Models \\ with one qubit}


\author{Ben O`Neill}{
  address={The School of Physics and Astronomy, University of Leeds,
  Leeds, LS2 9JT, United Kingdom}}

\author{Tom M. Barlow}{
  address={The School of Physics and Astronomy, University of Leeds,
  Leeds, LS2 9JT, United Kingdom}}

\author{Dominik {\v S}afr{\' a}nek}{
  address={Czech Technical University in Prague, B{\v r}ehov{\' a} 1, 
  115 19 Praha 1, Czech Republic}}

\author{Almut Beige}{
  address={The School of Physics and Astronomy, University of Leeds,
  Leeds, LS2 9JT, United Kingdom}}

\begin{abstract}
Hidden Markov Models (HMMs) have become very popular as a computational tool for the analysis of sequential data. They are memoryless machines which transition from one internal state to another, while producing symbols. These symbols constitute the output of the machine and form an infinite time series. Analogously, Hidden Quantum Markov Models (HQMM) produce an infinite time series, while progressing from one quantum state to another through stochastic quantum operations. Here we compare 1-bit HMMs and 1-qubit HQMMs and show that the latter can produce stronger correlations, although both machines are, in principle, comparable in resources.
\end{abstract}

\maketitle

\section{Introduction}

Hidden Markov Models (HMMs) are useful computational tools for the simulation of stochastic processes \cite{HMMs,Xue,Bart}. They are stochastic finite-state generators which can be used to model a wide range of situations. Applications include speech recognition, image analysis, and the modelling of biological systems. This paper concerns their quantum mechanical counterparts \cite{Wiesner,Monras,Gmeiner}. We especially discuss Hidden Quantum Markov Models (HQMMs) which have recently been introduced by Monras {\em et al.}~\cite{Monras}. Analogously to HMMs, we expect HQMMs to find a wide range of applications, including the simulation of quantum systems and the modelling of biological processes \cite{bio}. As shown in Ref.~\cite{Monras}, HQMMs naturally encompass HMMs in the sense that they are able to simulate the same and even more complex output sequences with the same number of internal states.

In the next section, we present a widely used and relatively intuitive definition of HMMs. Afterwards, we introduce HQMMs by following the ideas of Ref.~\cite{Monras}. For simplicity, we focus our attention on 1-bit and 1-qubit versions of these machines and assume output alphabets which contain only two symbols. These specific machines can be characterised relatively easily by a countable set of parameters which allows us to study their properties analytically and numerically. Calculating for example the maximum probabilities for certain words to occur, one can show that 1-qubit HQMMs are able to produce stronger correlations than 1-bit HMMs, although requiring, in principle, a similar amount of resources. 

\section{Hidden Markov Models}

Hidden Markov Models (HMMs) are machines which transition from one state of a finite set ${\cal S}$ to another state, while producing a symbol of an output alphabet ${\cal A}$ \cite{HMMs,Xue,Bart}. Although the actual state $Y \in {\cal S}$ of the machine remains hidden, it fully determines the probability for the system to progress to a certain state $X \in {\cal S}$ and to generate a certain output symbol $i \in {\cal A}$ in the next time step (cf.~Fig.~1.a). Different definitions of HMMs can be found in the literature. The HMMs which we consider here are known as Mealy (or transition-emitting) HMMs \cite{Xue}. Mealy HMMs encompass Moore (or state-emitting) HMMs, i.e.~they contain them as a subset \cite{Bart}.

For simplicity, we consider in the following 1-bit versions of transition-emitting HMMs. These produce either a 0 or a 1, while evolving between two states $A$ and $B$. Concrete versions of 1-bit HMMs are fully characterised by eight conditional probabilities $P(X,i|Y)$. Given these probabilities, the probability to obtain the output symbol $i$ at $t+1$ equals $P_{t+1}(i) = \sum_{X=A,B} \sum_{Y=A,B} P(X,i|Y) \, P_t(Y)$, if the machine is in state $Y$ at time $t$ with probability $P_t(Y)$. Analogously, one can show that the probability to transition to $X$ at $t+1$ equals $P_{t+1}(X) = \sum_{i = 0,1} \sum_{Y=A,B} P(X,i|Y) \, P_t(Y).$ Introducing the stochastic matrices $T_0$ and $T_1$ such that 
\begin{eqnarray} \label{Ti}
T_i &\equiv & \left( \begin{array}{cc} P(A,i|A) & P(A,i|B) \\ P(B,i|A) & P(B,i|B) \end{array} \right) \, , 
\end{eqnarray}
the above probability $P_{t+1}(X) $ can be written in a more compact way. Combining all of the above equations, one can show that the probability to find for example the word 10001 with the first letter being produced at $t+1$ equals
\begin{eqnarray}  
P_{t+1,...,t+5}(10001) &=& \left( \begin{array}{cc} 1 & 1 \end{array} \right) \, T_1 T_0 T_0 T_0 T_1 \, \left( \begin{array}{c} P_t(A) \\ P_t(B)  \end{array} \right) \, ,
\end{eqnarray}
if the machine is at time $t$ with probability $P_t(Y)$ in $Y$. Moreover, one can show that the machine is at $t+1$ in the statistical mixture described by
\begin{eqnarray}  \label{Tsum}
\left( \begin{array}{c} P_{t+1}(A) \\ P_{t+1}(B)  \end{array} \right) &=& \left( T_0 + T_1 \right) \left( \begin{array}{c} P_t(A) \\ P_t(B)  \end{array} \right)  \, ,
\end{eqnarray}
when the respective output symbols are disregarded. 

\begin{figure} \label{fig1}
  \includegraphics[height=.13 \textheight]{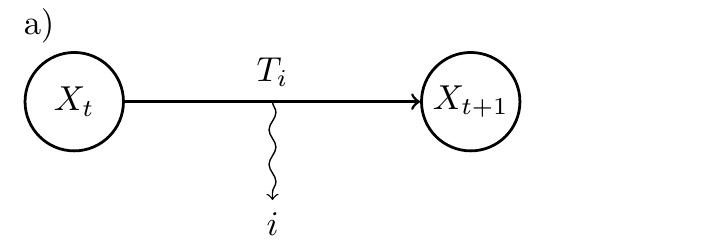} 
  \includegraphics[height=.13 \textheight]{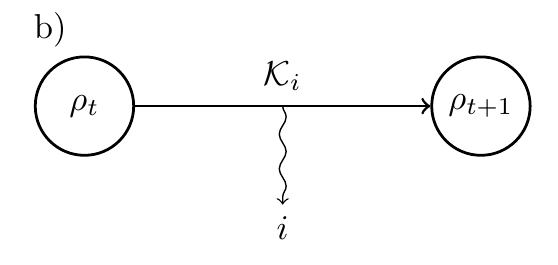} \\[0.4cm]
  \caption{Schematic view of the time evolution of a) HMMs and b) HQMMs.}
\end{figure}

In summary, a Mealy HMM is a stochastic finite-state generator \cite{Wiesner} which is characterised by a tuple $\{{\cal S}, {\cal A}, {\bf T} \}$, where ${\cal S}$ denotes a finite set of states, ${\cal A} = \{ i \}$ is a countable alphabet of output symbols, while ${\bf T} = \{ T_i : i \in {\cal A} \}$ is a set of square substochastic matrices of the same dimension as ${\cal S}$ \cite{Wiesner}. The machine undergoes transitions from one internal state to another. Each transition has an associated output symbol. The matrices $T_i$ summarise the probabilities for certain transitions to occur and fully characterise the machine. As illustrated in Eq.~(\ref{Tsum}), the sum $T = \sum_{i \in {\cal A}} T_i$ is a stochastic matrix which describes a Markovian evolution of the state space, when the output symbols are disregarded. 

\section{Hidden Quantum Markov Models}

Let us now have a closer look at possible quantum versions of HMMs. For example, a quantum finite-state generator \cite{Wiesner} which produces a similar output of 0's and 1's as a 1-bit HMM consists of a single qubit. The qubit evolves in time according to a unitary operator $U$ and undergoes projective measurements. A straightforward generalisation of a quantum finite-state generator is a 1-qubit machine which experiences generalised quantum measurements \cite{Kraus}. These are, for example, described by two Kraus operators $K_0$ and $K_1$ with
\begin{eqnarray} \label{condi}
K_0^\dagger K_0 + K_1^\dagger K_1 &=& 1 \, . 
\end{eqnarray}
Suppose the qubit is at time $t$ in a pure state $|\psi_t \rangle$. Then it changes at $t+1$ into $|\psi_{t+1} \rangle = K_i |\psi_t \rangle /\| \cdot \|$, if the output symbol $i \in \{ 0,1\}$ is detected. The corresponding output probability is $\|K_i |\psi_t \rangle \|^2$. One way of implementing $K_0$ and $K_1$ in the laboratory is to entangle the qubit with an ancilla qubit. The ancilla subsequently undergoes a projective measurement which produces either a 0 or a 1. Before the next time step, the ancilla should be resetted, if we want to realise a memory-less machines.

In order to obtain 1-qubit machines which cannot be outperformed by 1-bit HMMs, we need to consider the most general transformation which a qubit can undergo \cite{Monras}. The general definition of Hidden Quantum Markov Models (HQMMs) by Monras {\em et al.}  \cite{Monras} hence involves complete sets of Kraus operators ${\cal K}$. Each HQMM is characterised by a set $\{{\cal H},{\cal A},{\cal K}\}$, where ${\cal H}$ is the Hilbert space of a quantum system, ${\cal A}$ is again a countable alphabet of output symbols, and ${\cal K} =  \{K_i^m : i \in {\cal A}, 1 \ge m \ge M_i \}$ is a complete set of Kraus operators \cite{Kraus} such that
\begin{eqnarray}  \label{condi2}
\sum_{i \in {\cal A}} \sum_{m=1}^{M_i} K_i^{m\dagger} K_i^m &=& 1 \, .
\end{eqnarray}
In each time step, an output symbol $i \in {\cal A}$ is generated which depends only on the input state $\rho_t$ of the machine. If the output symbol $i \in {\cal A}$ is recorded, then the machine transitions into $\sum_{m=1}^{M_i} K_i^m \rho_t K_i^{m\dagger} / {\rm Tr} (\cdot)$. The corresponding output probability $P_{t+1}(i)$ equals $\sum_{m=1}^{M_i} {\rm Tr}(K_i^m \rho_t K_i^{m\dagger})$. If the output is ignored, the density matrix of the machine becomes 
\begin{eqnarray}
\rho_{t+1} &=& \sum_{i \in {\cal A}} \sum_{m=1}^{M_i} K_i^m \rho_t K_i^{m \dagger} 
\end{eqnarray}
at $t+1$, as illustrated in Fig.~1.b. For example, one way of realising a 1-qubit HQMM is to select a set of Kraus operators $\{ K_0, K_1\}$ which fulfills Eq.~(\ref{condi}) with a certain classical probability and to apply the corresponding Kraus operation ${\cal K}$ to the qubit. This means, HQMMs have been defined such that they can exploit quantum coherences as well as classical stochastic processes to generate relatively complex time series.

\section{Comparison}

The time evolution of HMMs and HQMMs is Markovian. This means, the states and output symbols of these machines depend only on the respective state of the machine in the previous time step. However, a 1-qubit HQMM can be prepared in a huge variety of states, while a 1-bit HMM is always either in $A$ or $B$. The state of a 1-qubit HQMM hence depends in general in a more complex way on the history of the machine. We therefore expect that HQMMs are able to produce more complex time series than HMMs, even when using a similar amount of resources.

It has already been shown that HQMMs encompass HMMs \cite{Monras}. This means, they are able to simulate the same output sequences as HMMs with the same number of internal states. However, it is not known how much more efficient HQMMs are in producing long-range correlations. To show that HQMMs are indeed more efficient, we recently determined the maximum probability for HQMMs and for HMMs to generate words of the form 100...0001 as a function of the words length. As reported in Ref.~\cite{Ben}, we found that 1-qubit HQMMs are able to produce stronger correlations than 1-bit HMMs. However, a systematic study of the efficiency and the information theoretic and physical properties of HQMMs is still outstanding.

\begin{theacknowledgments}
T.~B.~acknowledges financial support from a White Rose Studentship Network. 
\end{theacknowledgments}

\bibliographystyle{aipproc}

\end{document}